\documentclass[sigconf,10pt,twocolumn]{acmart}

\usepackage{mathtools}
\usepackage{tikz}
\usepackage{tikz}
\usepackage{listings}
\usepackage{comment}
\usepackage{enumerate}
\usepackage{color}
\usepackage{relsize}
\usepackage{booktabs}
\usepackage{graphics}
\usepackage{multirow}

\usepackage[toc,page]{appendix} 

\usepackage{halloweenmath}
\usepackage{subcaption}
\usepackage{amsthm}
\usepackage{amsfonts}
\usepackage[ruled,linesnumbered]{algorithm2e}
\newcommand*{\defeq}{\stackrel{\text{def}}{=}}
\let\oldnl\nl% Store \nl in \oldnl
\newcommand{\nonl}{\renewcommand{\nl}{\let\nl\oldnl}}% Remove line number for one line
\SetKwInput{KwData}{Input}
\SetKwInput{KwResult}{Output}
\DontPrintSemicolon

\usepackage{xcolor}
\usepackage{url}
\usepackage{booktabs}
\newtheorem{proposition}{Proposition}
\newtheorem{definition}{Definition}
\newtheorem{lemma}{Lemma}
\usepackage{mathtools}

\definecolor{codegreen}{rgb}{0,0.6,0}
\definecolor{codegray}{rgb}{0.5,0.5,0.5}
\definecolor{codepurple}{rgb}{0.58,0,0.82}
\definecolor{backcolour}{rgb}{0.95,0.95,0.92}
\usepackage{listings}
\lstdefinestyle{mystyle}{
    backgroundcolor=\color{backcolour},   
    commentstyle=\color{codegreen},
    keywordstyle=\color{magenta},
    numberstyle=\tiny\color{codegray},
    stringstyle=\color{codepurple},
    basicstyle=\ttfamily\footnotesize,
    breakatwhitespace=false,         
    breaklines=true,                 
    captionpos=b,                    
    keepspaces=true,                 
    numbers=left,                    
    numbersep=5pt,                  
    showspaces=false,                
    showstringspaces=false,
    showtabs=false,                  
    tabsize=2
}

\theoremstyle{definition}
\usepackage[utf8]{inputenc}

\newcommand{\figscale}{1.02}

\begin{document}

% \include{sigmod23_revision_request}
% \newpage
% \clearpage
% \setcounter{page}{1}

\title{Semantically Secure Private Set Intersection over Outsourced Multi-Owner Secret-Shared Databases}

% \author{Olamide Timothy Tawose}
% \affiliation{%
%   \institution{University of Nevada, Reno}
%   \country{United States}}
% \email{otawose@nevada.unr.edu}

% \author{Jun Dai}
% \affiliation{%
%   \institution{California State University, Sacramento}
%   \country{United States}}
% \email{jun.dai@csus.edu}

% \author{Lei Yang}
% \affiliation{%
%   \institution{University of Nevada, Reno}
%   \country{United States}}
% \email{leiy@unr.edu}

\author{Dongfang Zhao}
\affiliation{%
  \institution{University of Nevada, Reno}
  \country{United States}}
\email{dzhao@unr.edu}

\begin{abstract}
Private set intersection (PSI) aims to allow users to find out the commonly shared items among the users without revealing other membership information.
The most recently proposed approach to PSI in the database community was Prism,
which is built upon secret sharing and the assumption that multiple non-colluding servers are available.
One limitation of Prism lies in its semantic security:
the encoding on the servers is deterministic,
implying that the scheme cannot be indistinguishable under a chosen-plaintext attack (IND-CPA).
This paper extends the original PSI scheme of Prism by two orthogonal primitives,
namely Kaleido-RND and Kaleido-AES:
the former exhibits highly efficient performance with randomized encoding and the latter is provably secure under CPA attacks with more computational overhead.
A system prototype is implemented and deployed on a 34-node cluster of SQLite instances.
Extensive experiments on the TPC-H benchmark and three real-world applications confirm the effectiveness of the proposed Kaleido primitives.
\end{abstract}

\settopmatter{printfolios=true}
\maketitle

\section{Introduction}

Private set intersection (PSI) aims to allow users to find out the commonly shared items among the users without revealing other membership information.
As can be used as a versatile cryptographic primitive, 
PSI has been actively studied in the literature of applied cryptography~\cite{bpinkas_eurocrypt20} and recently in privacy-preserving outsourced databases~\cite{yli_sigmod21}.
As a concrete example,
PSI has been recognized as a crucial building block to support various operations in outsourced databases,
such as semi-joins~\cite{fkers_sac12}.

The most recently proposed approach to PSI in the database community was Prism~\cite{yli_sigmod21},
which is built upon secret sharing and the assumption that multiple non-colluding servers are available.
Secret sharing~\cite{ashamir_cacm79}, as the name suggests,
allows users to split a plaintext into multiple portions such that a single portion (stored on a distinct server) does not reveal any information about the plaintext.
and is usually considered as a specific form of secure multiparty computation (MPC)~\cite{ayao_focs82}.
Prism demonstrates a high efficiency that is based on group-homomorphic exponentiation.

One limitation of Prism~\cite{yli_sigmod21} lies in its semantic security:
Although the exact number of shared entities is masked through number-theoretical modular exponentiation,
such encoding is deterministic,
implying that the scheme cannot be indistinguishable under a chosen-plaintext attack (IND-CPA).
The first goal of this paper is,
therefore,
to describe an attack that can reveal information more than the intersection among parties. 

To guarantee the semantic security of Prism,
the second goal of this paper is to extend the original PSI scheme of Prism by two orthogonal primitives:
(i) \textit{Kaleido-RND}, an efficient module to randomize the group generator based on which the exponentiation is computed in the codomain of a multiplicative group,
and (ii) \textit{Kaleido-AES}, a pseudorandom function to ensure the IND-CPA security of the entire PSI scheme.
As the name suggests, our proposed schemes aim to garble the original Prism scheme,
ending up a scattered prism---looking like a kaleidoscope.
While Kaleido-RND is highly efficient and does offer randomization on the ciphertext,
Kaleido-RND is not proven IND-CPA secure.
However, we will demonstrate that Kaleido-AES is provably semantically secure with the cost of computational overhead.

To demonstrate the effectiveness of the proposed Kaleido schemes,
we implement a system prototype with about 1,000 lines of Python and Shell script and deploy the system on a 34-node cluster of SQLite instances on CloudLab~\cite{cloudlab}.
We evaluate Prism, Kaleido-RND, and Kaleido-AES with the TPC-H benchmark and three real-world applications,
where we scale the number of database owners (i.e., clients) between 2 and 32.
Our results show that Kaleido-RND only incurs insignificant overhead on the server encoding and is generally negligible in the entire system.
Moreover, although Kaleido-AES seems to introduce much computational overhead compared to the original Prism,
the overhead is orders of magnitude lower than the communication cost,
demonstrating the practicality of Kaleido-AES in addition to its provable security.

In summary, this paper makes the following technical contributions.
\begin{itemize}
    \item We demonstrate a security vulnerability of a state-of-the-art PSI scheme, Prism, in outsourced databases. We provide both the intuition behind an effective attack and detail such an attack with concrete examples.~(\S\ref{sec:prism})
    
    \item We propose a series of new primitives, namely Kaleido, to extend Prism such that the new PSI scheme is semantically secure.
    We prove the IND-CPA security of Kaleido under the widely-accepted computational assumption and exemplify its correctness with both theoretical analysis and case studies.~(\S\ref{sec:kaleido})
    
    \item We implement the Kaleido scheme with about 1,000 lines of Python and Shell scripts.
    We deploy the system on two servers and 32 database owners hosted in the public cloud, CloudLab~\cite{cloudlab}.
    Extensive experiments on the TPC-H benchmark and three real-world applications confirm the effectiveness of the proposed Kaleido schemes.~(\S\ref{sec:eval})
\end{itemize}

\section{Preliminary}

\subsection{Private Set Intersection and Secure Multi-Party Computation}

Private set intersection (PSI)~\cite{mfreedman_eurocrypt04,yhuang_ndss12,skamara_fc14,kvladimir_ccs16,bpinkas_sec15,bpinkas_tps18,bpinkas_eurocrypt20} has been extensively studied in outsourced databases,
such as being used as a building block for privacy-preserving joins between outsourced databases~\cite{jmoha_ccs20,sbad_ccs22}.
Indeed, PSI can be considered a special form of set operations among multiple parties,
and therefore can be implemented through a general-purpose secure multi-party computation scheme,
which mostly reply on arithmetic or boolean gates that incur significant computation overhead.

Secure multiparty computation (MPC)~\cite{ogold_stoc87,ylindell_cacm21,daran_latincrypt21} has a long history~\cite{ayao_focs82}.
The goal of MPC is more ambitious:
in addition to keeping the plaintext confidential,
we want to calculate an arbitrary function of the original plaintexts by touching on only the encoded data on multiple parties.
The original problem was solved by the so-called garbled circuits~\cite{ayao_focs82},
whose idea was pretty simple:
we can ask each party to encode the input with its private key,
shuffle the encrypted ciphertexts,
and then enumerate all the keys to decrypt the result.
Since we assume the encryption scheme is secure,
the only way that the result can be revealed is that the correct combination of private keys is applied to one of the garbled outputs.
This is indeed a feasible solution, at least theoretically;
in practice, the circuits may grow exponentially and result in efficiency issues.
There are many more efficient MPC solutions,
such as~\cite{dbeav_stoc90,mnaor_ec99,vkole_asiacrypt05,szahu_eurocrypt15}.

One of the main limitations of PSI schemes lies in their scalability.
In fact, many existing PSI schemes support only two parties,
such as~\cite{mion_sp20,bpin_eurocrypt18,hchen_ccs17}.
More recent works~\cite{jcheon_tfeccs12,chaz_pkc17,rinbar_scn18,lkiss_crypto05,vkol_ccs17,ple_ccs19,yli_sigmod21} emerged to focus efficient PSI schemes on more than two parties.
This work falls into the category of the latter:
The proposed Kaleido scheme supports an arbitrary of parties for the PSI operation.

\subsection{Secret Sharing}

The idea of a secret sharing scheme (SSS) is straightforward:
a given plaintext $pt$ is converted into a set of encoded bytes $ct$'s such that only a specific \textit{subset} of $ct$'s can reconstruct the original $pt$.
The goal of SSS is to reduce the risk of disclosing the plaintext;
instead of compromising the holder of the plaintext,
the malicious adversary needs to subvert multiple entities before any of the shareholders detect the attack.
Even for weaker attacks where only semi-honest adversaries are assumed, 
dispersing the secret shares to more parties raises the bar of a successful eavesdropping attack.

In practice, a SSS can be tuned by the subset size. 
Formally, a $(t,n)$-threshold SSS (TSSS) is defined as follows.
\begin{definition}[TSSS]
A $(t,n)$-TSSS is comprised of two algorithms:
\begin{itemize}
    \item Share: a randomized algorithm that takes as input a plaintext $pt$ and returns a sequence $S = (s_1, \dots, s_n)$ of shares.
    \item Reconstruct: a deterministic algorithm that takes a set of at least $t$ shares and returns the plaintext.
\end{itemize}
The number $t$ is called the threshold of the TSSS.
Let $U$ of size $t$ be a subset of $n$ shares, $|U| \ge t$ and $U \subseteq \{s_1, \dots, s_n\}$,
we require that a TSSS holds the following property:
\[
Reconstruct(U) = pt.
\]
\end{definition}
As we will see in the next section~\S\ref{sec:provable}, the definition of TSSS leads to a slightly different security definition compared with the conventional encryption schemes.

The canonical example of $(t,n)$-TSSS is due to Shamir~\cite{sham_ccam79},
in which the secrets were revealed through a $(t-1)$-degree polynomial.
In essence, each share can reconstruct the coefficient of a specific degree of unknowns through the LaGrange polynomials.
In addition to Shamir's construction, 
other schemes exist.
Ito et al.~\cite{mito_ecj89} proposed the replicated secret-sharing scheme,
which was based on finite fields where each share is a vector.
One nice property of replicated secret-sharing is its linearity:
the addition and subtraction of local shares can be linearly transformed into the addition and subtraction of the plaintexts.
A simpler variant of replicated secret-sharing is \textit{additive secret sharing},
where each share is a scalar value and the threshold $t$ is set to $n$.

\subsection{Probably Security}\label{sec:provable}

When employing an encryption scheme in an application,
it is highly desirable to demonstrate its security provably.
Formally, we need to identify the following three important pieces for the provable security of a given encryption scheme:
security goal, threat model, and assumption.
The security goal spells out the desired effect when the application is under attack;
the threat model articulates what an adversary can do with the attack,
such as what information of the plaintext/ciphertext can be collected and the resource/time limitation of the attack;
the assumption lists the presumed conditions of the cryptographic scheme (e.g., factoring a product of two big primes, finding the discrete logarithmic root).
The security goal and threat model are usually called \textit{security definition} collectively.

One well-accepted security definition with a good balance between efficiency and security is that the adversary can launch a \textit{chosen-plaintext attack} (CPA),
defined as follows.
\begin{definition}[Chosen-Plaintext Attack]
Given a security parameter $n$,
i.e., the bitstring length of the key,
an adversary can obtain up to $poly(n)$ of plaintext-ciphertext pairs $(m, c)$,
where $m$ is arbitrarily chosen by the adversary and $poly(\cdot)$ is a polynomial function in $n$.
With such information, the adversary tries to decrypt a $c'$ that is not included in the polynomial number of known ciphertexts.
\end{definition}

The polynomial requirement mandates that the adversary should only be able to run a polynomial algorithm without unlimited resources.
Accordingly, we want to design encryption schemes that are \textit{CPA secure}: 
even if the adversary $\mathcal{A}$ can obtain those extra pieces of information, 
$\mathcal{A}$ should not be able to decode the ciphertext better than a random guess up to a very small probability.
To quantify the degree of this small probability,
\textit{negligible function} is defined as below.

\begin{definition}
A function $\mu(\cdot)$ is called negligible if for all polynomials $poly(n)$ the inequality $\mu(n) < \frac{1}{poly(n)}$ holds for sufficiently large $n$'s.
\end{definition}

For completeness, we list the following lemmas for negligible functions that will be used in later sections.
We state them without proof, 
which can be found in introductory cryptography or complexity theory texts.

\begin{lemma}[Arithmetic operation between two negligible functions is a negligible function]
\label{thm:neg_sum}
Let $\mu_1(n)$ and $\mu_2(n)$ be both negligible functions.
Then the following functions are all negligible:
$\mu_1(n) + \mu_2(n)$, $\mu_1(n) - \mu_2(n)$, $\mu_1(n) \times \mu_2(n)$, and $\frac{\mu_1(n)}{ \mu_2(n)}$.
\end{lemma}

The canonical method to prove the security of a proposed encryption scheme,
such as IND-CPA,
is through \textit{reduction}~\cite{ylind_book17}.
Usually, breaking the scheme is \textit{reduced} to solving a hard mathematical problem,
which means that if an attack is possible for the scheme then the mathematical problem would be efficiently solved.
That is, the encryption scheme is at least as hard as the mathematical problem.
The scheme is modeled as a subroutine,
whose inputs are simulated such that the adversary cannot tell whether it is being involved in an attack or in a subroutine to help solve the hard problem. 
Although forward proof is possible,
the more commonly used technique is a contradiction:
by assuming that the adversary could distinguish two ciphertexts in an experiment with a non-negligible advantage, 
the reduction would lead to a non-negligible probably to efficiently solve the hard mathematical problem that is believed to be intractable,
thus leading to a contradiction.

\section{\texttt{Prism} PSI}
\label{sec:prism}

This section discusses the potential security issues of Prism~\cite{yli_sigmod21}.
We start by describing the Prism PSI scheme~(\S\ref{sec:prism_scheme}),
then provide a security analysis~(\S\ref{sec:prism_analysis}),
and finally, exemplify the security issues using a real-world example~(\S\ref{sec:prism_example}).

\subsection{Scheme Definition of Prism PSI}
\label{sec:prism_scheme}

This subsection provides a brief introduction to Prism~\cite{yli_sigmod21}.

\subsubsection{Assumptions}

\paragraph{Non-colluding Servers}
Prism assumes that a cluster of non-colluding servers is available for storing secret shares generated from the original plaintext.
In practice, this assumption can be realized by, for example, acquiring virtual machines from distinct cloud computing vendors such as Google Cloud Platform and Amazon Web Services.

\paragraph{Hardness of Discrete Logarithmic Problem}
Let $\mathbb{G}$ denote a (multiplicative) cyclic group and $g$ as a generator.
That is, $\forall y \in \mathbb{G}$, $\exists x \in [0, |\mathbb{G}| - 1]$ such that $y = g^x$.
Informally, the discrete logarithmic (DL) problem states that if the carnality of $\mathbb{G}$ is sufficiently large,
then it is infeasible to find $x$ in polynomial time even if both $y$ and $g$ are given.
This is a well-accepted assumption in most modern cryptographic schemes,
such as ECDSA~\cite{ecdsa} and ElGamal~\cite{elgamal_tit85}.

\paragraph{Knowledge of Clients}
We assume that the clients are aware of the domain of the attribute on which the PSI is carried on.
In the context of PSI,
we are mostly interested in the categorical or discrete data~\cite{yli_sigmod21},
implying that each client can represent its local items with a boolean vector,
where each element is a boolean value indicating whether a specific item exists on the client.

\subsubsection{Architecture}

While the full-fledged Prism deployment comprises four subsystems:
clients (i.e., database owners), non-colluding servers, the oracle (initiator), and the announcer,
the PSI functionality can be completed through the first three subsystems, i.e., without the announcer.

\paragraph{Clients (Database Owners)}
The clients are the owners of the sensitive data that should be kept confidential from other clients.
We expect the clients are \textit{honest} during the PSI procedure:
(i) clients do not have the incentive to lie about their local data items for PSI and are not interested in probing or analyzing any intermediate results;
(ii) clients are not compromised by adversaries without being detected, 
which can be enforced by digital signatures~\cite{ecdsa} and Byzantine fault-tolerant protocols~\cite{castro_pbft02}.

\paragraph{Servers}
In the PSI scheme of Prism, two servers are available and they do not collude. 
This implies that the servers cannot be active adversaries;
that is, both servers are not malicious. 
However, servers are not trusted in the sense that they may behave as passive adversaries,
meaning that they could be interested in learning about the plaintexts from its involvement in the outsourced databases.
Because the data touched on by each of the two servers are random due to the protocol of generating secret shares,
each server has no way to probe the plaintext and is thus information-theoretically secure.

\paragraph{Oracle}
The oracle is also called an \textit{initiator},
whose job is to set up the initial parameters for both the clients and the servers.
For example, two groups are required for Prism and it is the oracle that specifies the group order and generator.
The oracle is useful only at the beginning of the PSI scheme;
if the context is clear, we will skip the oracle's initialization step and assume it has been done in the offline stage.
It should be noted that the \textit{oracle} in Prism is not a conventional oracle that has been widely used in cryptography:
in the latter case, 
an oracle usually refers to an entity whose functionality is ``ideal'' and does not reveal any internal machinery, i.e., as a black box.

\begin{figure}[!t]
  \centering
  \includegraphics[width=0.9\linewidth]{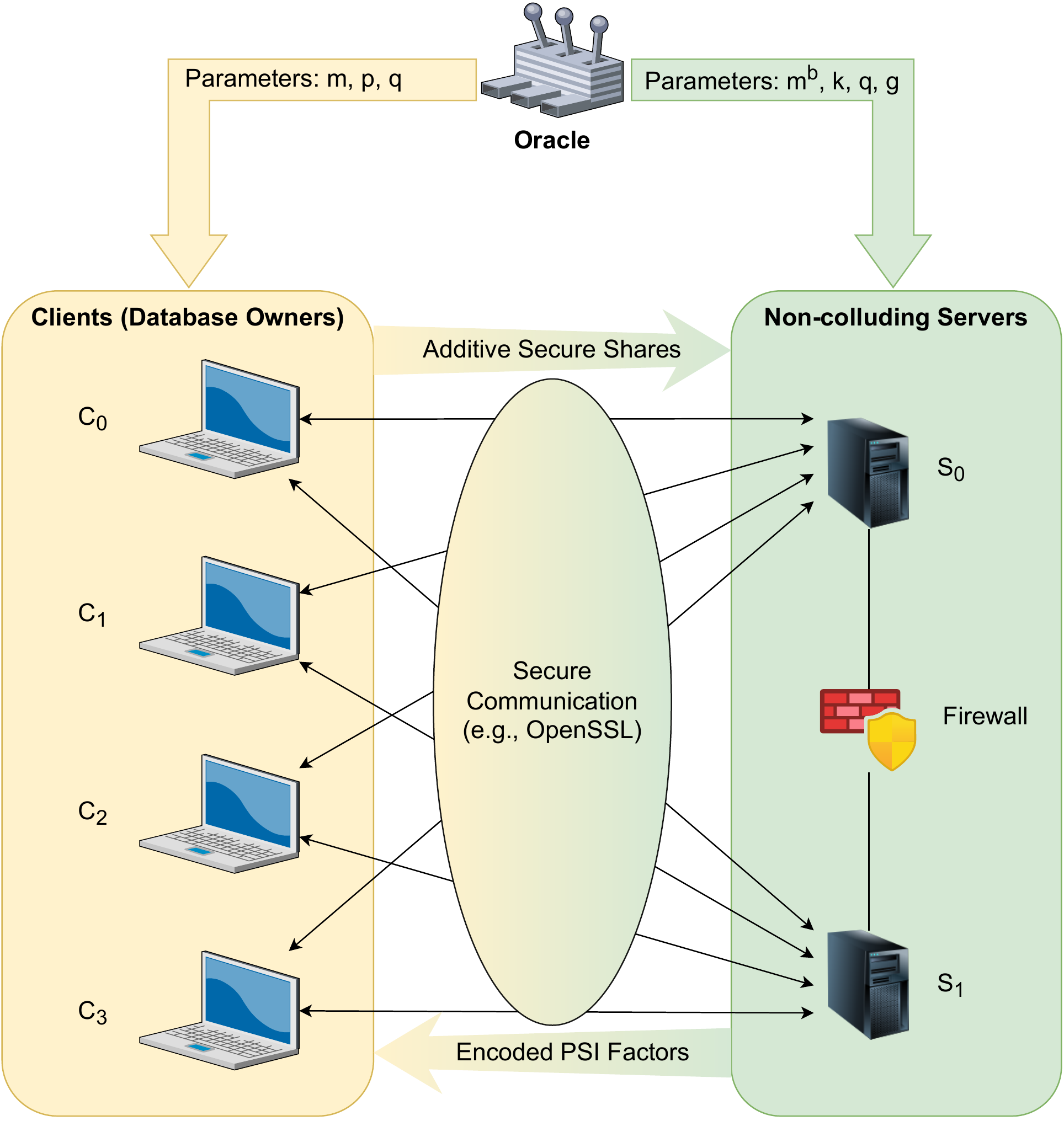}
  \caption{System model of PSI over multi-owner databases}
  \label{fig:MOD_PSI_Arch}
\end{figure}

Figure~\ref{fig:MOD_PSI_Arch} illustrate the system model of PSI over multi-owner outsourced databases.
We assume there are four clients $\{C_0, C_1, C_2, C_3\}$ and two non-colluding servers $\{S_0, S_1\}$.
Before the clients start outsourcing the local data,
the oracle sends parameters to clients and servers.
The clients then apply additive secret-sharing primitives to the local data and send them to both servers through a secure communication channel, such as OpenSSL~\cite{openssl_github}.
Both servers execute a protocol to encode the secret shares and broadcast the encoded values back to all clients.

\subsubsection{Protocols}
\label{sec:prism_protocol}

\paragraph{Initialization}
The oracle picks the parameters of two cyclic groups and broadcasts them to servers and/or clients.
Some of those parameters (e.g., generator of the multiplicative group) should be kept confidential to the servers only such that the clients cannot learn about any information beyond the PSI.
In addition to the parameters, the oracle is also responsible for configuring the platform such as the host names and other metadata information.

\paragraph{Vectorization of Attribute Domain}
Each client $C_i$ generates a local vector,
$V_i$, such that $\forall v \in V_i \implies v \in \{0, 1\}$.
A value $v = 1$ in $V_i$ if and only if the corresponding item exists on $C_i$;
otherwise $v = 0$.
Indeed, here we assume that all clients agree upon the same order of items in the domain of the attribute, denoted by $A_c$.
On straightforward approach to agreeing on the order of domain values is to take the alphanumeric order.

\paragraph{Secret Sharing}
Each client $C_i$ splits its local vector $V_i$ into two secret vectors, denoted by $V_i^0$ and $V_i^1$ that are to be sent to $S_0$ and $S_1$, respectively.
One requirement for the splitting is that the secrets must be \textit{additive},
meaning that $V_i^b$, $b \in \{0, 1\}$, must be able to reconstruct $V_i$ by the addition operation on the secrets.
The additive property exploited by Prism is the modular addition defined in an additive group $\mathbb{Z}_p = \{0, 1, \dots, p-1\}$, 
where $p$ denotes a large prime.
That is, given a plaintext $v \in \{0, 1\}$\footnote{Recall that the elements of a vector are all boolean values, 0 or 1.}, 
we randomly pick the first share $v^0 \in [0, m)$ and calculate the second share $v^1$ as follows:
\[
v^1 \equiv v - v^0 \text{ (mod p)}.
\]
If we apply the above splitting to all elements of a local vector $V_i$,
we can generalize the above equation from scalars to vectors,
in an element-wise fashion,
as the following:
\[
V_i^1 \equiv V_i - V_i^0 \text { (mod p)}.
\]

\paragraph{Aggregating Shares}
After server $S_b$ receives all the secret shares $V_i^b$,
$b \in \{0, 1\}$ and $0 \le i < m$,
$S_b$ carries out a local aggregation over the secret shares from all clients $C_i$.
That is, server $S_b$ computes a new vector $V^b$ as follows:
\begin{equation}
\label{eq:aggr_server}
V^b \equiv -[m^b] + \sum_{i = 0}^{m-1} V_i^b \text{ (mod p)},
\end{equation}
where $[m^b]$ denotes the extended vector (with the same length of $V^b$) of a secret share generated from the number of clients $m$, $b \in \{0, 1\}$,
in the same way we previously compute for $v^b$.
The operand of $-m^b$ is added because we want to in later steps cancel out the value ``1'' in a vector that represents the existence of a shared attribute value.

\paragraph{Encoding Shares}
The server should not simply send the locally aggregated shares back to the client because doing so would reveal the exact number of shared items among clients.
Instead, the servers must somehow encode the partial shares such that the client cannot learn about the shared items except for the intersection.
The encoding scheme used by Prism unsurprisingly leverages the algebraic property of a homomorphic function $Hom$ between two cyclic groups $\mathbb{G} = (G, +)$ and $\mathbb{H} = (H, \times)$ such that:
\begin{equation}\label{eq:hom}
    \begin{split}
        Hom: G & \to H \\
        x & \mapsto g^x \text{ (mod q)}, 
    \end{split}
\end{equation}
where $g$ is a generator and $q$ is the order of the multiplicative group $\mathbb{H}$.
It is obvious that the function $Hom()$ is homomorphic,
$\forall x_0, x_1 \in G$:
\[
Hom(x_0) \times Hom(x_1) = g^{x_0} \times g^{x_1} = g^{x_0 + x_1} = Hom(x_0 + x_1).
\]
We can then apply $Hom()$ to the elements of $V^b$ on server $S_b$.
Let $U^b$ denote the encoded vector in which each element is applied with the $Hom()$ function.
Server $S_b$ broadcasts $U^b$ to all clients.

\paragraph{Calculating PSI}
Each client receives two vectors $U^0$ and $U^1$ from two servers $S_0$ and $S_1$, respectively.
The client calculates the element-wise product of $U^0$ and $U^1$ without knowing the value of $g$---the generator of the multiplicative group.
Let $n$ denote the cardinality of $A_c$'s domain;
if $U^0 = (u^0_0, u^0_1, \dots, u^0_{n-1})$ and $U^1 = (u^1_0, u^1_1, \dots, u^1_{n-1})$,
then we define the element-wise product as follows:
\begin{equation}\label{eq:ctxt}
\displaystyle
U = U^0 \odot U^1 \defeq \left(u^0_0 \cdot u^1_0, u^0_1 \cdot u^1_1, \dots, u^0_{n-1} \cdot u^1_{n-1}\right).
\end{equation}
Evidently, $\forall i \in [0, n)$, $u_i \defeq u^0_i \cdot u^1_i = 1$ if and only if the summation of secret shares of all clients equals $m$. 
Indeed, this is the criterion for detecting the intersection of the attribute values:
only those values whose encoding is 1 from all $m$ clients should be included in the PSI.
On the other hand, if not all boolean ciphertexts are 1,
then $u_i$ cannot be 1 implying that the $i$-th value of the attribute domain is not in PSI.
Furthermore, because $g$ is visible only to the servers,
clients cannot learn about the exact number of clients who share the $i$-th value of the attribute domain.

\subsection{Security Analysis of Prism}
\label{sec:prism_analysis}

\subsubsection{High-Level Intuition}

While Prism hides the exact number of clients who share the $i$-th value of the attribute domain unless every client holds it,
this section will show that there is other information leaked from the Prism scheme.
The key insight is that the function of Eq.~\eqref{eq:hom} is \textit{deterministic},
implying that repeated inputs lead to the same output.
As a result, if the same number of clients share a specific attribute value,
then Prism would leak such information to an adversary.

\subsubsection{Chosen-Plaintext Attacks}

\paragraph{Security Notion}
Informally, a \textit{Chosen-Plaintext Attack} (CPA) refers to an attack where the adversary can distinguish the ciphertexts of two arbitrarily chosen plaintexts with a non-negligible probability,
even after running the attack algorithms polynomial times in the security parameter, usually the bitstring length of the problem size, e.g., the cardinality of the attribute domain.
A \textit{negligible function} of argument $n$ is defined as 
\begin{equation}\label{eq:negl}
\displaystyle
negl(n) \defeq o(n^{-c}),
\end{equation}
where $c$ denotes any constant.
That is equivalent to:
\[\displaystyle
\forall c > 0 \implies \lim_{n \to \infty} \frac{negl(n)}{n^{-c}} = 0.
\]
The formal proof of \textit{indistinguishability under CPA} (IND-CPA) for a scheme $\Pi$ usually takes a form of a simulation,
where we assume that $\Pi$ is not IND-CPA and use $\Pi$ to construct a protocol that would solve an intractable problem,
thus leading to a contradiction.
In the context of PSI, IND-CPA implies that all elements in $U$ should look random except for 1,
which indicates a shared attribute value.
An adversary can trivially break the IND-CPA of Prism in two rounds:
\begin{itemize}
    \item The adversary inserts a new attribute value $x_n$ to all but the first client and retrieves the corresponding element in $U$, say $u_n$, from any client.
    \item The adversary inserts a new attribute value $x_{n+1}$ to all but the first client and another new attribute value $x_{n+2}$ to all but the first two clients.
    Similarly to the first round, the adversary retrieves two ciphertexts in $U$: $u_{n+1}$ and $u_{n+2}$.
\end{itemize}
The adversary can win the IND-CPA game with a probability of one by comparing $u_n$ with $u_{n+1}$ and $u_{n+2}$.
While it can be argued that the practicality of allowing the adversary to insert new attribute values into database owners is debatable,
the CPA security does consider the ``worst-case'' scenario.
Even if this IND-CPA may be thought of as theoretical interest only,
the following section constructs a more practical attack.

\subsubsection{Inference beyond PSI}

Suppose that each client computes the encoded vector $U = (u_0, u_1, \dots, u_{n-1})$ from $C_i$, $i \in [0, m)$,
where $n$ denotes the cardinality of the attribute domain.
Without loss of generality, let $0 \le i \not= j < n$ and $u_i = u_j \not= 1$.
This means the $i$-th and the $j$-th values of attribute $A_c$ are shared by up to $m-1$ clients.
Let $[\cdot]$ denote the positional element of an array,
a specific client $C_k$ can infer the following information:
\begin{itemize}
    \item If $V_k[i] = V_k[j]$, $C_k$ knows that $A_c[i]$ and $A_c[j]$ are equally ``popular'' on all database owners. 
    \item If $V_k[i] < V_k[j]$, i.e., $V_k[i] = 0$ and $V_k[j] = 1$, 
    then $C_k$ knows that there is exactly one additional client who holds $A_c[i]$ rather than $A_c[j]$.
    Depending on the applications,
    such information might introduce bias or unfairness,
    e.g., in a market of competitors.
    \item If $V_k[i] > V_k[j]$, client $C_k$ can infer similar information that is symmetric to the second case above.
\end{itemize}
Evidently, any of the above three scenarios does reveal some information beyond the intersection of items among the clients.

\subsection{CPA Attack on Prism}
\label{sec:prism_example}

This section demonstrates a running example of Prism's PSI protocol and attacks on it.

\subsubsection{System Configuration}
We assume $m = 4$,
i.e., there are four clients, $C_i$, $0 \le i < 4$.
If not otherwise stated, let $b \gets \{0, 1\}$,
i.e., $b$ is assigned 0 or 1 with 50\% probability\footnote{In the literature of cryptography, it is also common to use $\overset{\$}{\gets}$, $\gets_R$, or other varieties to denote the same thing.}.
There are a fixed number of non-colluding servers;
for simplicity, two servers are available: $S_0$ and $S_1$.
Let the additive group $\mathbb{G} = (\mathbb{Z}_5, +)$ and the multiplicative group $\mathbb{H} = (\mathbb{Z}_{11}^*, \times)$ and $g=3$ as a generator of a subgroup of $\mathbb{H}$.
Note that the aforementioned parameters are taken for the sake of explanation;
they are usually much larger in practice to be resilient to brute-force attacks.

\subsubsection{Schema and Data}

\paragraph{Schema and Metadata}
For simplicity, we assume that all of the four clients hold their local relations of a single-attribute schema.
Since there is only a single attribute, 
we can safely represent the column of the relationship as a vector.
Let's also assume the data type of the single attribute is an integer in the semi-open interval [0, 5).
We use $n$ to denote the cardinality of the attribute domain,
i.e., $n = 5$ in our example.
This means that although each client could store an arbitrary number of values,
the length of each plaintext vector is 5, 
i.e., $|V_i| = 5$, $0 \le i < 4$.

\paragraph{Data}
Let $R_i$ denote the single-attribute relation on client $C_i$ as a list:
\begin{equation}
\label{eq:rawdata}
\begin{cases}
R_0 = (0, 1, 3) \\
R_1 = (1, 3, 4) \\
R_2 = (3, 4, 4) \\
R_3 = (1, 2, 3, 4)
\end{cases}
\end{equation}
The goal of PSI is for every $C_i$ to find out the commonly shared item,
in this case integer 3.

\subsubsection{Vectorization}
The plaintext data on four clients in Eq.~\eqref{eq:rawdata} can be vectorized to the following:
\begin{equation}
\label{eq:vector}
\begin{cases}
V_0 = (1, 1, 0, 1, 0) \\
V_1 = (0, 1, 0, 1, 1) \\
V_2 = (0, 0, 0, 1, 1) \\
V_3 = (0, 1, 1, 1, 1) 
\end{cases}
\end{equation}
The elements that should be included in the PSI are those values whose positional vector elements are one.
In our example, the only element satisfying this is integer 3.
If we denote the PSI as a function over the set of relations, 
we have
\[
PSI(\mathcal{R}) = \{3\},
\]
where $\mathcal{R} \defeq \{R_i\}$, $0 \le i < m$.
We also introduce another parameterized function $Card(\cdot, k)$,
which returns the set of attribute values that are shared by $k$ clients:
\[\displaystyle
\begin{split}
    Card: R^m \times \mathbb{Z}_m & \to 2^{A_c}\\
    \left(\mathcal{R}, k\right)& \mapsto \{y \;|\; y \subseteq A_c\},
\end{split}
\]
where $R$ denotes an arbitrary vector of integers between 0 and $n$ and $R^m$ denotes the cross-product $m$ times over $R$.
For simplicity, we sometimes place the second parameter in the subscript,
i.e., $Card_k(\cdot) \defeq Card(\cdot, k)$.
By definition, it is evident that $PSI(\mathcal{R})$ is a special case of $Card_k(\mathcal{R})$,
i.e., $PSI(\mathcal{R}) \equiv Card_{|\mathcal{R}|}(\mathcal{R})$.
With this notation, 
we can write the following for our example:
\begin{equation}
\label{eq:card}
\begin{cases}
Card_0(\mathcal{R}) = \emptyset\\
Card_1(\mathcal{R}) = \{0, 2\}\\
Card_2(\mathcal{R}) = \emptyset\\
Card_3(\mathcal{R}) = \{1, 4\}\\
Card_4(\mathcal{R}) = \{3\}\\
\end{cases}
\end{equation}
We notice that the cardinality of some $Card_k()$ is equal,
such as $|Card_1(\mathcal{R})| = |Card_3(\mathcal{R})| = 2$.
As we will see later, this will reveal information that can be exploited by the adversary.

\subsubsection{Secret Sharing}

Recall that the client relies on the additive group $\mathbb{G} = (\mathbb{Z}_5, +)$ for additive secret sharing, 
as discussed in~\S\ref{sec:prism_protocol}.
Therefore, the $V$'s in Eq.~\eqref{eq:vector} can be split into the following:
\[
\begin{cases}
V_0^0 = (3, 4, 1, 2, 0); \; V_0^1 = (3, 2, 4, 4, 0)\\
V_1^0 = (1, 2, 4, 3, 4); \; V_1^1 = (4, 4, 1, 3, 2)\\
V_2^0 = (0, 4, 2, 3, 1); \; V_2^1 = (0, 1, 3, 3, 0)\\
V_3^0 = (2, 3, 4, 1, 0); \; V_3^1 = (3, 3, 2, 0, 1)
\end{cases}
\]
It can verified that $V_i = V_i^0 + V_i^1$ in $\mathbb{G}$, $0 \le i < 4$.
Client $C_i$ then sends $V_i^0$ to server $S_0$ and $V_i^1$ to server $S_1$.

\subsubsection{Server Encoding}
Each of the two servers receives $m = 4$ vectors of shares.
That is, server $S_b$ receives $V_i^b$, $b \in \{0, 1\}$, $0 \le i < m$.
As specified in Eq.~\eqref{eq:aggr_server},
$m$ should be split as well on $\mathbb{G}$.
In this example, let $m^0 = 1$ and $m^1 = 3$.
Therefore, the first element of the aggregated vector $V^0$ on $S_0$ can be calculated as
\[
V^0[0] = -m^0 + \sum_{i=0}^3 V_i^0 = -1 + (3 + 1 + 0 + 2) \equiv 0 \text{ (mod 5)}.
\]
Similarly, we can compute both aggregated vectors
\[
\begin{cases}
    V^0 = (0, 2, 0, 3, 4)\\
    V^1 = (2, 2, 2, 2, 0)
\end{cases}
\]
The two servers then apply Eq.~\eqref{eq:hom} to $V^b$ and generated two new vectors (note that $q = 11$ in our example):
\[
\begin{cases}
    U^0 = (1, 9, 1, 5, 4)\\
    U^1 = (9, 9, 9, 9, 1)
\end{cases}
\]
Both servers then broadcast $U^b$ to all clients.

\subsubsection{Client Decryption}

All clients receive two vectors $U^b$ from both servers.
Each client computes the element-wise multiplication of $U^b$ under group $\mathbb{H}$, i.e., modular 11.
For example,
\[
U^0[1] \times U^1[1] = 9 \times 9 = 81 \equiv 4 \text{ (mod 11)},
\]
and 
\[
U^0[3] \times U^1[3] = 5 \times 9 = 45 \equiv 1 \text{ (mod 11)}.
\]
The final vector on each client is therefore 
\[
U = (9, 4, 9, 1, 4).
\]
All clients at this point know that the fourth value of $A_c$ is shared by all clients because $U[3] = 1$ and have no idea about how many other clients hold what other values because those encrypted values are meaningless due to Eq.~\eqref{eq:hom}.

\subsubsection{Information Leakage}
All clients learn that $U[0] = U[2] = 9$.
Let's see what client $C_0$ can learn from this.
Recall that client $C_0$ knows its local vector,
e.g., $V_0[0] = 1$ and $V_0[2] = 0$.
This means that $C_0$ can infer the following fact:
there must be at least one client $C_j$, $j \not= 0$, 
such that $V_j[2] = 1$,
because otherwise $U[0]$ cannot equal $U[2]$.
That is, by following the Prism PSI scheme $C_0$ learns that the integer ``2'' exists on at least one of the other three clients, although ``2'' is not shared by all of the four clients (because $U[2] = 9 \not= 1$).

\section{\texttt{Kaleido} PSI}
\label{sec:kaleido}

To fix the information leakage of Prism PSI, 
we design a new PSI scheme called Kaleido for multi-owner databases.
As the name suggests, Kaleido introduces disorder to the PSI calculation among multi-owner databases and therefore is considered ``garbled'' Prism.
The key idea of Kaleido is fairly simple:
instead of using a fixed generator,
let's randomize it such that the encoded shares on the server cannot be distinguished even for the same inputs.
The technical challenge of applying this idea is to ensure both the correctness and the security of PSI.

\subsection{Server Protocol of Kaleido}
\label{sec:kaleido_protocol}

\paragraph{Description}
We elaborate on the randomness introduced by Kaleido servers in Algorithm~\ref{alg:kaleido}.
Lines 1 -- 7 simply aggregate the secret shares sent by all clients and subtract the secret share of the number of clients $m$.
Line 10 ensures that 
(i) the output of the parameterized function is indistinguishable from a random string, and
(ii) the output is identical on both $S_0$ and $S_1$ because the secret key $k$ is agreed upon between both servers.
In practice, $PRF_k(\cdot)$ can be implemented by a block cipher,
such as AES~\cite{aes}.
Kaleido then maps the randomized output to an element in the group $\mathbb{H}$,
which is done in Line 11.
As in Line 10, both $S_0$ and $S_1$ generate the same $g$ because $\mathbb{H}$ is revealed to both servers.
Lines 12--14 ensure that $g$ is a generator of a (sub)group of order $p$.
Note that we use $|g|$ (Line 12) to denote the \textit{order} of $g$ in the cyclic group rather than the absolute value of scalar $g$ in arithmetic.
On Line 15, the elevated value in group $\mathbb{H}$ replaces the element in the group $\mathbb{G}$.
Line 16 ensures that even the same elements in $\mathbb{G}$ map to distinct elements in $\mathbb{H}$ due to different values of $ct$ and $g$.
Server $S_b$ finally broadcasts its local encoding $V^b$ to all clients,
as shown in Line 18.

\begin{algorithm}
\SetAlgorithmName{Algorithm}{}{}
\caption{\textit{Enc( )}: Server Protocol of Kaleido PSI}\label{alg:kaleido}
\KwData{
Two servers $S_0$ and $S_1$;
$m$ clients $C_i$, $0 \le i < m$, each sending $V^b_i$ to $S_b$, $b \in \{0, 1\}$;
the length of vector $V^b_i$ is $n$;
oracle $\mathcal{O}$ splits $m$ into $m^b$ such that $m \equiv m^0 + m^1 \text{ (mod p)}$ and sends $m^b$ to $S_b$;
a shared key $k$ that is agreed upon by $S_b$;
a keyed pseudorandom function $PRF(\cdot)$;
two cyclic groups $\mathbb{G} = (\mathbb{Z}_p, +)$ and $\mathbb{H} = (\mathbb{Z}^*_q, \times)$;
initial vector (IV) generated by $\mathcal{O}$;
}
\KwResult{
$S_b$ output $V^b$ such that (i) $V^0 \odot V^1$ has no identical elements except for ``1'', and (ii) element equals ``1'' if and only if all $m$ clients encode the element as ``1'' (cf. Eq.~\eqref{eq:vector});
}

\nonl \;
\nonl $/**$ \;
\nonl $\hspace{2.5mm}*$ \hspace{5mm} Aggregate Secret Shares \;
\nonl $\hspace{2.5mm}*/$ \;
$V^b = Numpy.zeros((1,n))$ \;
\For{i = 0; i < n; i++}{
    \For{j = 0; j < m; j++}{
        $V^b[i] \coloneqq V^b[i] + V^b_j[i]$\;        
    }
    $V^b[i] \coloneqq V^b[i] - m^b \;\%\; p$\;
}

\nonl \;
\nonl $/**$\;
\nonl $\hspace{2.5mm}*$ \hspace{5mm} Encode Aggregated Shares\;
\nonl $\hspace{2.5mm}*/$ \;
$pos \coloneqq 0$\;
\For{each $v \in V^b$}{
    $ct \gets PRF_k(pos \oplus IV)$\;
    $g \coloneqq (ct \;\%\; (q - 2)) + 2$\tcp*{[2, q-1]}
    \While{$|g| \not= p$}{
        $g \coloneqq ((g + 1) \;\%\; (q - 2)) + 2$\tcp*{$g$ generator}
    }
    $V^b[pos] \coloneqq g^v$ \% $q$\;
    $pos$++\;
}

\nonl \;
\nonl $/**$\;
\nonl $\hspace{2.5mm}*$ \hspace{5mm} Return Ciphertext Vector\;
\nonl $\hspace{2.5mm}*/$ \;
$S_b$ broadcasts $V^b$ to $C_i$, $0 \le i < m$\;

\end{algorithm}

\begin{figure}[!t]
  \centering
  \includegraphics[width=0.9\linewidth]{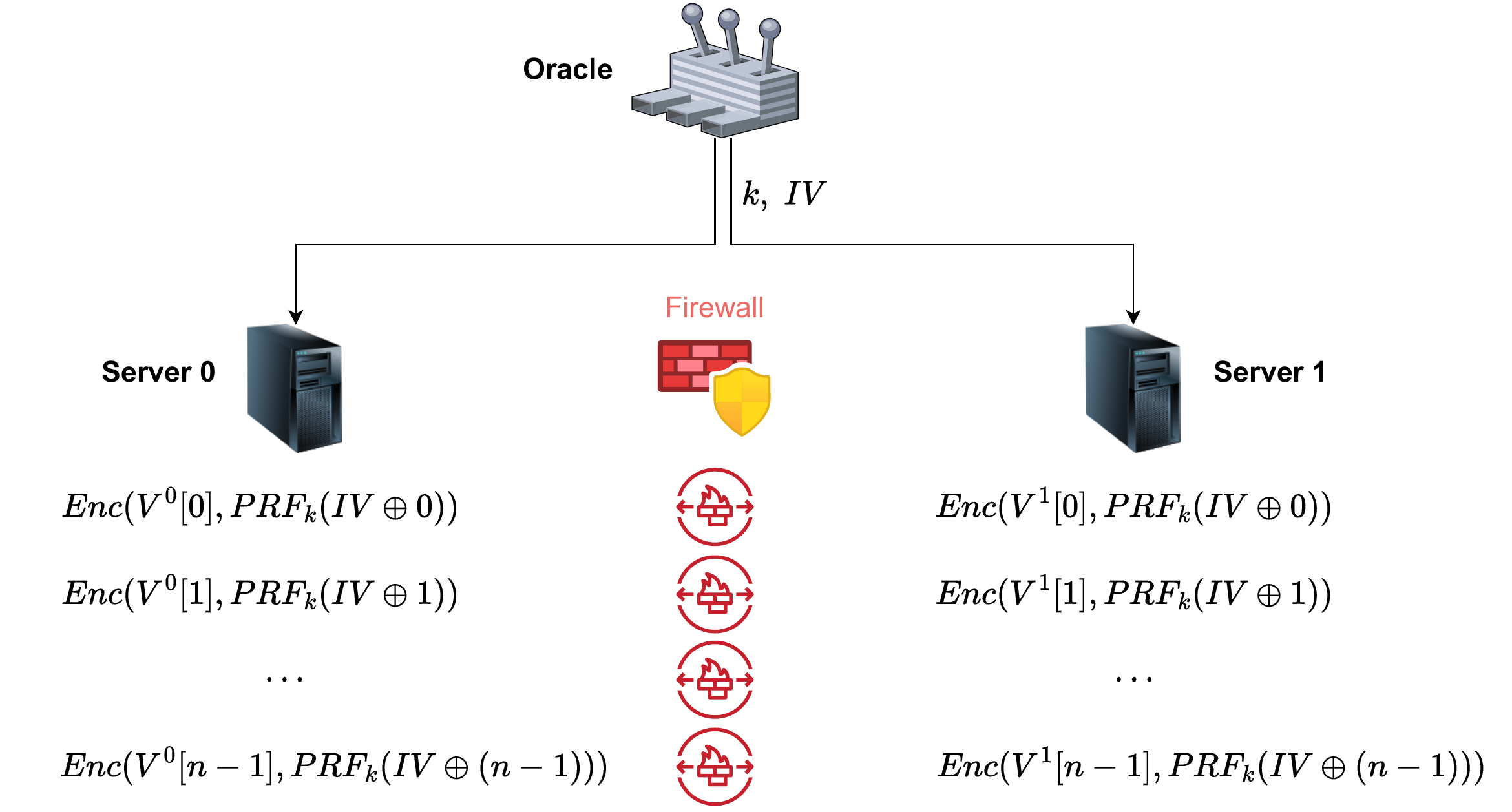}
  \caption{Random ciphertext generated by two non-colluding servers in Kaleido PSI}
  \label{fig:Kaleido}
\end{figure}

Figure~\ref{fig:Kaleido} illustrates a simple example where both servers take an initial vector (IV) to randomly encode the secret shares submitted by clients.
Intuitively, because both servers parameterize the pseudorandom function $PRF(\cdot)$ with the same key $k$ initialized by the oracle at the beginning of the execution,
the base number $g$ (cf. Line 11, Alg.~\ref{alg:kaleido}) is identical on both servers and yet look random in different rounds.

\paragraph{Complexity}
Suppose the cost of computing $PRF(\cdot)$ is $c$.
Lines 2 -- 7 take $\mathcal{O}(nm)$.
Lines 9 -- 17 take $\mathcal{O}(nqc)$.
Therefore, the overall computational complexity of Alg.~\ref{alg:kaleido} is $\mathcal{O}(nm+nqc)$.
The overall number of messages is straightforward:
Line 18 incurs $m$ messages.
Since there are two servers,
the overall number of network messages is $2m$.
The overall number of communication rounds is 1,
also incurred by Line 18 (assuming \textit{broadcasting} is implemented as an asynchronous primitive).

\subsection{Correctness of Kaleido}

We need to show that $U[i] = 1$, $0 \le i < n$, if and only if $V_j[i] = 1$, $0 \le j < m$.

\subsubsection{$V_j[i] = 1 \implies U[i] = 1$}

If $V_j[i] = 1$,
recall that this means the $j$-th client holds the $i$-th positional value of attribute $A_c$.
Let $g_i$ denote the $i$-th base of elevation,
as shown in Line 12 of Alg.~\ref{alg:kaleido}.
As discussed in~\S\ref{sec:kaleido_protocol},
$g_i$ is identical for both $S_0$ and $S_1$.
According to Eq.~\eqref{eq:ctxt}, the following holds
\begin{equation}\label{eq:correctness}
\displaystyle
\begin{split}
U[i]    & \equiv U^0[i] \times U^1[i] \text{ (mod q)}\\
        & = g_i^{V^0[i]} \times g_i^{V^1[i]}\\
        & \equiv g_i^{V^0[i] + V^1[i] \text{ (mod p)}}\\
        & = g_i^{ - m^0 - m^1 + \sum_{j=0}^{m-1} V_j[i]}\\
        & = g_i^{ - m + \sum_{j=0}^{m-1} 1}\\
        & = g_i^0\\
        & = 1.
\end{split}
\end{equation}

\subsubsection{$U[i] = 1 \implies V_j[i] = 1$}

If $U[i] = 1$, then according to Eq.~\eqref{eq:correctness},
the following equality must hold:
\[
- m^0 - m^1 + \sum_{j=0}^{m-1} V_j[i] = 0,
\]
or equivalently:
\[
\sum_{j=0}^{m-1} V_j[i] = m.
\]
Recall that $V_j[i] \in \{0, 1\}$, cf. Eq.~\eqref{eq:vector}.
Therefore, the only way to satisfy the above equation is $V_j[i] = 1$, $0 \le j <m$,
which is desired.

\subsection{Provable Security of Kaleido}

We claim that the PSI scheme backed by Alg.~\ref{alg:kaleido} is IND-CPA except for the attribute values shared by all clients.
This means that the adversary is not allowed to query the servers to return the ciphertext of an attribute value that is shared by all clients.
We formulate the above in the following proposition.

\begin{proposition}
If the $PRF(\cdot)$ function in Alg.~\ref{alg:kaleido} is indistinguishable from an ideal random number generator,
the elements of $U$ generated by $V^b$ in Alg.~\ref{alg:kaleido}, 
$b \in \{0, 1\}$, 
are provably secure in the IND-CPA security model.
\end{proposition}

\begin{proof}
Let $\mathcal{A}$ denote the adversary who can break the security of Alg.~\ref{alg:kaleido}.
We write $MOD^{CPA}_{\mathcal{A},\; Kaleido}$ to denote the private-set-intersection (PSI) experiment in Multi-Owner Databases (MOD):
\begin{enumerate}
    \item $\mathcal{A}$ outputs two attribute values $x_0$ and $x_1$ in $\mathbb{G}$ such that there exists at least one client that does not hold $x_0$ or $x_1$, respectively;
    \item Kaleido servers agree upon a value $b \in \{0, 1\}$, run Alg.~\ref{alg:kaleido}, and send $Enc(x_b)$ to $\mathcal{A}$;
    \item $\mathcal{A}$ outputs $b' \in \{0, 1\}$;
    \item The output of the experiment is defined to be 1 if $b' = b$, and 0 otherwise. We write $MOD^{CPA}_{\mathcal{A},\; Kaleido} = 1$ if the output of the experiment is 1 and $\mathcal{A}$ is said to be \textit{successful}.
\end{enumerate}
% It follows that the probability for $\mathcal{A}$ to succeed is not negligible:
% \[
% Pr[MOD^{CPA}_{\mathcal{A},\; Kaleido} = 1] > \frac{1}{2} + negl(\lambda),
% \]
% where $negl(\cdot)$ is a negligible function in the security parameter $\lambda$,
% as defined in Eq.~\eqref{eq:negl}.
Let $g_0$ and $g_1$ denote the bases of $x_0$ and $x_1$, 
respectively, as specified in Line 12 of Alg.~\ref{alg:kaleido}.
We consider two cases: (i) $g_0 \not= g_1$ and (ii) $g_0 = g_1$.
\begin{itemize}
    \item If $g_0 \not= g_1$, we have 
    \[\displaystyle
        Pr\left[MOD^{CPA}_{\mathcal{A},\; Kaleido} = 1 \;|\; g_0 \not= g_1\right] = \frac{1}{2},
    \]
    because the best $\mathcal{A}$ can do is a random guess of $b$.
    According to elementary probability theory, 
    we also know that the following holds:
    \begin{equation}\label{eq:cpa_1}
    \displaystyle
        Pr\left[MOD^{CPA}_{\mathcal{A},\; Kaleido} = 1 \wedge g_0 \not= g_1\right] \le \frac{1}{2}.
    \end{equation}
    
    \item If $g_0 = g_1$, since $PRF(\cdot)$ is indistinguishable from a random generator,
    the following equality holds:
    \[\displaystyle
    Pr\left[g_0 = g_1\right] = \frac{1}{2^\lambda}.
    \]
    It follows that 
    \begin{equation}\label{eq:cpa_2}
    \displaystyle
        Pr\left[MOD^{CPA}_{\mathcal{A},\; Kaleido} = 1 \wedge g_0 = g_1\right] \le \frac{1}{2^\lambda}.
    \end{equation}
\end{itemize}
    
Combining Eqs.~\eqref{eq:cpa_1} and~\eqref{eq:cpa_2},
we have
\[\displaystyle
Pr\left[MOD^{CPA}_{\mathcal{A},\; Kaleido} = 1\right] \le \frac{1}{2} + \frac{1}{2^\lambda} = \frac{1}{2} + negl(\lambda),
\]
implying that $\mathcal{A}$ cannot succeed significantly better than a random guess,
which completes the proof.
\end{proof}

\subsection{Case Study of Kaleido}
We revisit the example discussed before in~\S\ref{sec:prism_example} to illustrate the effectiveness of the proposed Kaleido PSI.
Recall that the probability of having $g_i = g_j$, $i \not= j$, 
is negligible.
Let's assume that $g_i$ equals $i+2 \;\%\; 11$.
For example, $U^0[3]$ is calculated as
\[
U^0[3] = (3+2 \;\%\; 11) ^ {V^0[3]} \;\%\; 11 = 5^3 \;\%\; 11 = 4,
\]
and $U^1[3]$ is calculated as
\[
U^1[3] = (3+2 \;\%\; 11) ^ {V^1[3]} \;\%\; 11 = 5^2 \;\%\; 11 = 3.
\]
Following the above algorithm, $U^b$ can be calculated as:
\[
\begin{cases}
U^0 = (1, 9, 1, 4, 9)\\
U^1 = (4, 9, 5, 3, 1)
\end{cases}
\]
Consequently, the Kaleido PSI $U$ is:
\[
U = (4, 4, 5, 1, 9).
\]
We see that $U[3] = 1$ still implies that the fourth positional value of $A_c$ is shared by all clients,
which is desired.
However, the equality or inequality between other values does not reveal the number of clients sharing the values.
For example, although $U[0] = U[1] = 4$,
it does not indicate that integers ``0'' and ``1'' are shared by the same number of clients;
in fact, ``0'' is hold only by $C_0$ and ``1'' is held by all clients but $C_2$.
Similarly, although $U[0] \not= U[2] = 5$, 
it does not mean that integers ``0'' and ``2'' are not shared by the same number of clients---both are indeed shared by the same number of clients (a single client $C_0$ for ``0'' and a single client $C_3$ for ``2'').

\section{Evaluation}
\label{sec:eval}

\subsection{System Implementation}

We have implemented the proposed Kaleido scheme with about 1,000 lines of Python code and Bash script,
which will be released at
\url{https://github.com/}.
We choose the lightweight SQLite as the local database instance.
Note that SQLite is a file-based database and does not support network access.
We thus implement a communication layer among remote SQLite instances through the \textit{paramiko} library for secure data transfer and remote query invocation.
Some of the most important libraries and dependencies include:
\textit{python} 3.8.0,
\textit{sqlite} 3.31.1,
\textit{numpy} 2.21.0,
\textit{paramiko} 2.12.0,
\textit{scp} 0.14.4, 
and \textit{cryptography} 39.9.0.

\subsection{Experimental Setup}

\subsubsection{Test Bed}

We deploy Kaleido and other baseline schemes alone with SQLite~\cite{sqlite} on a 34-node cluster hosted at CloudLab~\cite{cloudlab}.
Each node is equipped with two 32-core Intel Xeon Gold 6142 CPUs, 384 GB ECC DDR4-2666 memory, and 	
two 1~TB SSDs.
The operating system image is Ubuntu 20.04.3 LTS,
and the page size is 4 KB.
All servers are connected via a 1 Gbps control link (Dell D3048 switches) and a 10 Gbps experimental link (Dell S5048 switches).
We only use the experimental links for our evaluation.

Specifically, we name the 34 nodes in the cluster as \textit{node0}--\textit{node33}.
The two servers run on \textit{node0} and \textit{node1},
and the database owners (i.e., clients) are deployed on \textit{node2}--\textit{node33}.
All 34 nodes are enabled with password-less SSH connection for convenient communication since our evaluation focuses on performance metrics rather than security measurement. 
% Figure~\ref{fig:testbed} illustrates the star topology of our 34-node cluster.

% \begin{figure}[!t]
%   \centering
%   \includegraphics[width=.8\linewidth]{fig/testbed.png}
%   \caption{Kaleido topology on CloudLab~\cite{cloudlab}}
%   \label{fig:testbed}
% \end{figure}

\subsubsection{Systems under Comparison}

For all the systems under comparison, the orders of the additive and multiplicative groups are 113 and 227, respectively.

\paragraph{Prism}
Prism~\cite{yli_sigmod21} assigns a static generator during the server encoding.
As discussed in prior sections, 
Prism can leak information other than the PSI,
such as which items are shared by the same number of clients.
However, because the generator is fixed,
Prism incurs no extra cost for randomizing the ciphertext,
which leads to the highest performance compared with the schemes proposed in this paper.

\paragraph{Kaleido-RND (Naive Randomness)}
One straightforward way to resolve static generators in Prism is to randomly pick a generator,
which we refer to as \textit{Kaleido-RND}.
This requires that both servers synchronize the random seeds and therefore incur server-to-server communication,
which is against the assumption that the PSI servers should be non-colluding.
Moreover, we are unaware of an effective model to formally prove the security of this scheme.

\paragraph{Kaleido-AES (Provably Secure)}
We chose AES-128~\cite{aes} to implement the pseudorandom function in our Kaleido prototype system.
PKCS7 is used for padding the plaintext data and the cipher block chaining (CBC) mode is adopted for the AES cipher.
The initial vector (IV) is set to 1234.

\subsubsection{Workloads}

\paragraph{TPC-H Benchmark}
The synthetic data set is TPC-H version 3.0.0~\cite{tpch3}, a standard database benchmark.
There are overall eight tables in TPC-H;
in our evaluation, we select the LineItem table---the same one used in Prism~\cite{yli_sigmod21},
which comprises 6,001,215 tuples.
The attribute that we will focus on is \textit{L\_ORDERKEY},
which consists of 1,500,000 distinct values.
Each database owner, i.e., the client, 
randomly select 5,000,000 values from the \textit{L\_ORDERKEY} attribute,
and the PSI goal is to find out those \textit{L\_ORDERKEY} values shared by the set of clients.

\paragraph{COVID-19 Patient Record}
The first application is the U.S. national COVID-19 statistics from April 2020 to March 2021~\cite{covid19data}.
The data set has 341 days of 16 metrics, such as \textit{death increase}, \textit{positive increase}, and \textit{hospitalized increase}.
Each client randomly selects 300 days of data as their local data.

\paragraph{Bitcoin Trade History}
The second application is the history of Bitcoin trade volume~\cite{bitcoin_trade} since it was first exchanged in the public in February 2013.
The data consists of the accumulated Bitcoin exchange on a 3-day basis from February 2013 to January 2022,
totaling 1,086 large numbers.
Each client randomly selects 1,000 trade data points in their local database.
    
\paragraph{Human Gene \#38 Sequence}
The third application is the human genome reference 38~\cite{hg_data}, 
commonly known as \textit{hg38},
which includes 34,424 rows of singular attributes,
e.g., \textit{transcription positions}, \textit{coding regions}, and \textit{number of exons}, last updated in March 2020.
Each client randomly selects 30,000 rows in its local data set.
    
\subsection{Performance}

\subsubsection{TPC-H Benchmark}

\begin{figure}[!t]
  \centering
  \includegraphics[width=0.7\linewidth]{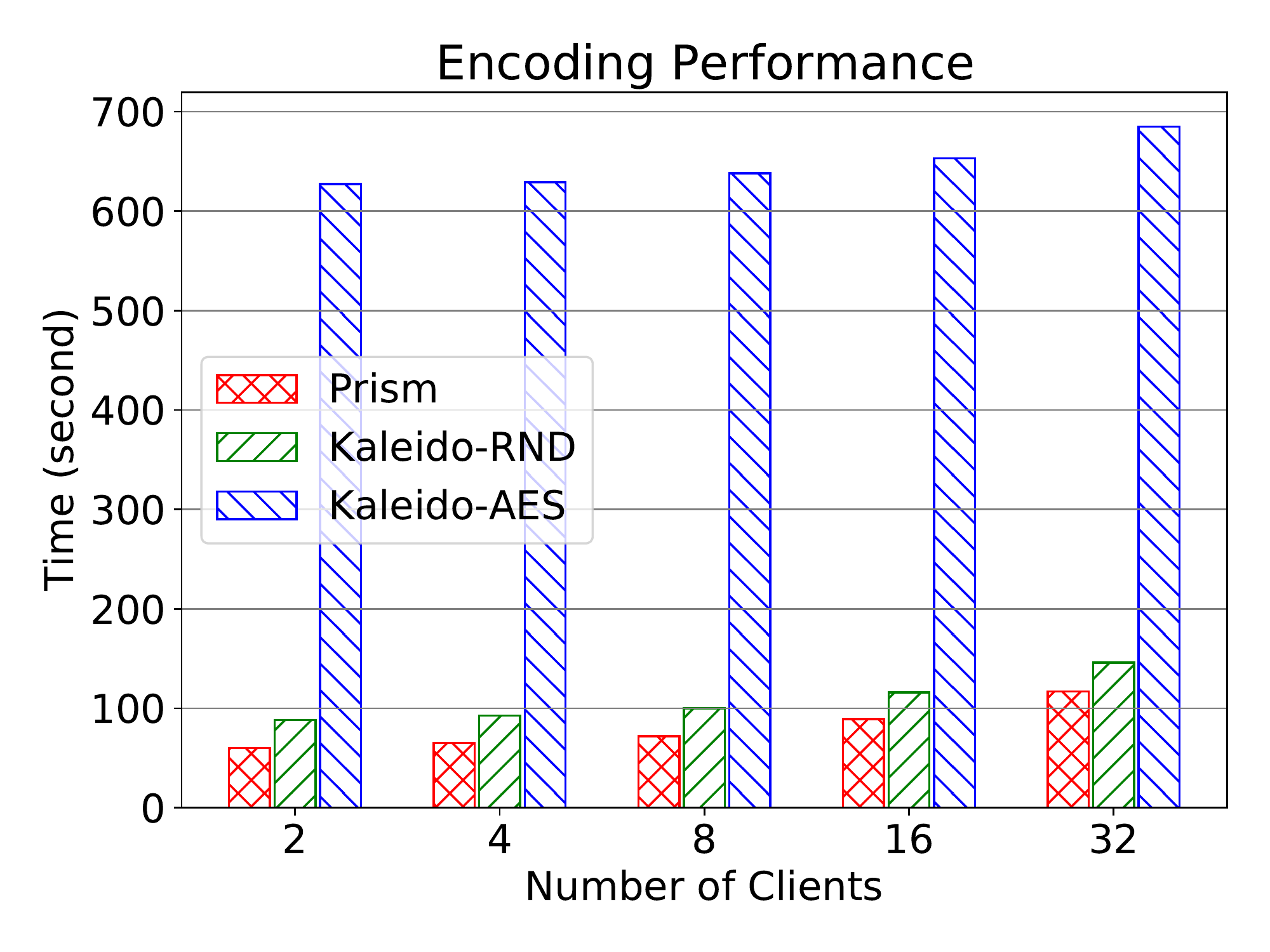}
  \caption{Encoding performance of various schemes}
  \label{fig:server}
\end{figure}

\paragraph{Server Encoding}
Figure~\ref{fig:server} reports the encoding performance of various schemes.
We compare the server processing time of all three schemes when serving different numbers of clients ranging from 2 to 32.
Unsurprisingly, Prism achieves the highest performance because of the fixed generator $g$.
Kaleido-RND is slightly slower than Prism because of the communication between the two servers and the randomization cost on both servers.
Kaleido-AES introduces about 500 seconds overhead for encoding five million tuples;
this is the cost for ensuring the semantic security of PSI over multi-owner databases.
Please note that the overhead is a computational cost,
which can be reduced by parallel processing that has not been explored in this work.

\paragraph{Client Computation}

In our implementation, the work on the client can be broken into four stages:
(i) Load the data from the local SQLite database,
(ii) Hash the raw value and map it to a binary flag,
(iii) Split the binary flag into two additive secret shares,
and (iv) Recover the PSI result from two encoded vectors from servers.
Table~\ref{tbl:compute_overhead} reports the time consumption for each of the above four stages.
We observe that the most costly stage is \textit{Split},
which takes almost 10 seconds.
Again, this can be parallelized with multiple CPU cores because the elements in the vector are independent,
which we will explore in our future work.

\begin{table}[t]
\caption{Computational time on clients (in seconds)}
\centering
\begin{tabular}{ c c c c c } 
\toprule
Benchmark & Load & Hash & Split & Recover\\
\midrule
TPC-H LineItem & 5.615 s & 2.682 s & 9.358 s & 0.139 s \\ 
\bottomrule
\end{tabular}
\label{tbl:compute_overhead}
\end{table}

% \begin{table}[h]
% \centering
% \caption{Communication cost, model size, number of values, and communication time for a single iteration on a single client}
% \label{tbl:client_compute}
% \begin{tabular}
% \toprule
% \textbf{Dataset} & \textbf{\small{Model Size (MB)}} & \textbf{\small{Number of Values}} & \textbf{\small{Communication Time (ms)}} \\
% \midrule
% CIFAR-10 & 0.061 & 53,761 & 0.488 \\
% FMNIST & 0.184 & 166,370 & 1.472 \\
% SVHN & 0.382 & 343,175 & 3.056 \\
% \bottomrule
% \end{tabular}
% \end{table}

\paragraph{I/O Cost}

Table~\ref{tbl:io_cost} reports the local I/O cost on both clients and servers.
Note that the numbers are for individual clients or servers,
e.g., each client persists a vector of 15 MB in size---the total I/O overhead of the entire system is thus $15 \times m$ MB, 
where $m$ denotes the number of database owners (clients).

\begin{table}[t]
\caption{I/O cost (in Megabytes)}
\centering
\begin{tabular}{ c c c } 
\toprule
Benchmark & Client Vector & Server Vector\\
\midrule
TPC-H LineItem & 15 MB & 17 MB \\ 
\bottomrule
\end{tabular}
\label{tbl:io_cost}
\end{table}

\begin{figure}[!t]
  \centering
  \includegraphics[width=0.7\linewidth]{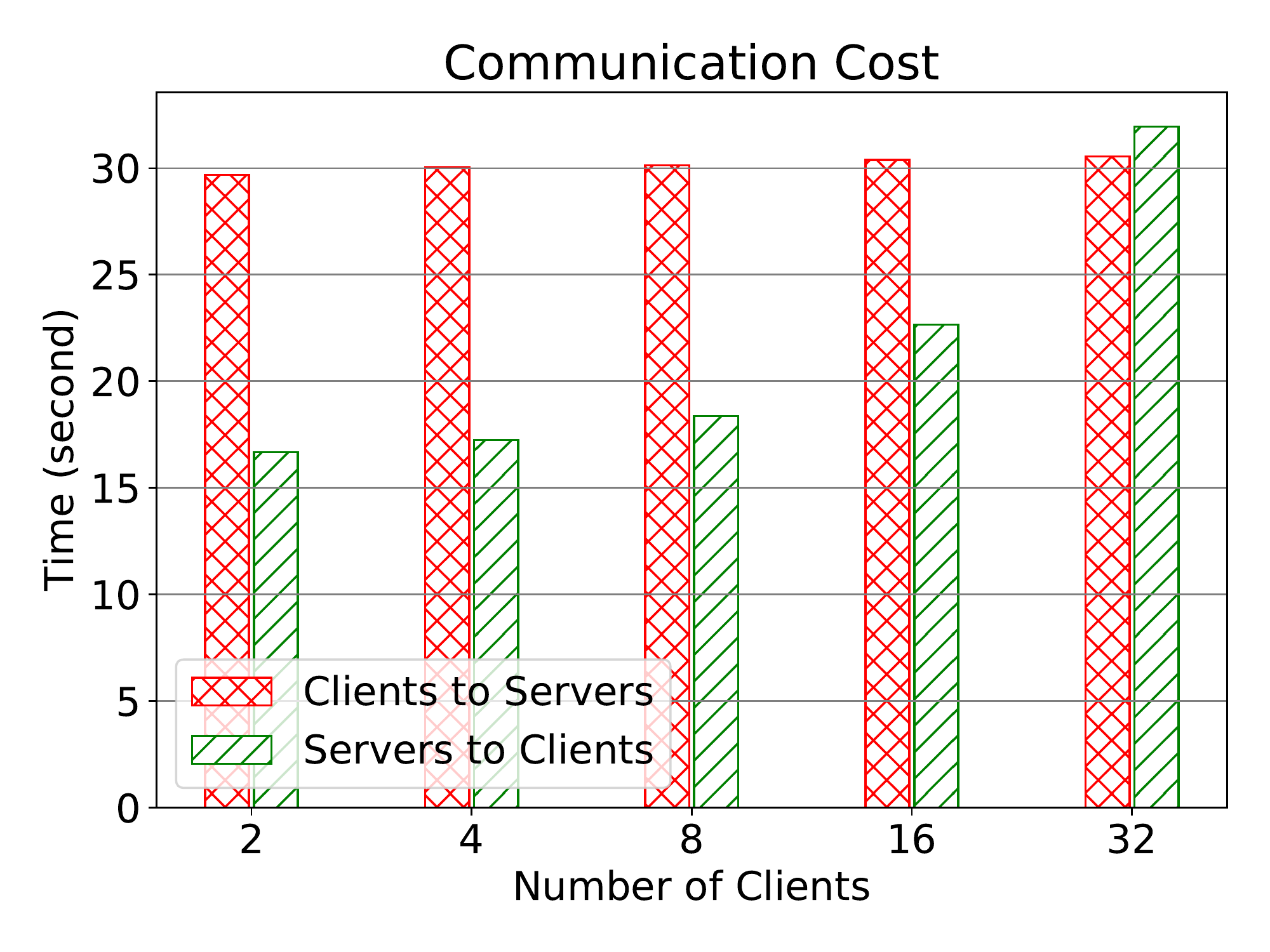}
  \caption{Communication time on different numbers of clients}
  \label{fig:comm}
\end{figure}

\paragraph{Communication Cost}
Figure~\ref{fig:comm} reports the communication time spent on different numbers of clients.
We observe that the upstream communication is almost constant when scaling out the number of clients;
however, the downstream traffic takes proportionally more time when more clients are involved.
When the number of clients is sufficiently large,
e.g., 32,
server-to-client communication takes more time than the other direction.

\subsubsection{Applications}

Due to limited space, we only report the performance of these three applications on client scales: 
two, four, and eight.
We observe that in these real-world applications,
the performance overhead lies in communication:
both the client and the server spend most of their time sending the vectors,
orders of magnitude more than other stages.

\paragraph{COVID-19}

\begin{figure}[!t]
  \centering
  \includegraphics[width=\figscale\linewidth]{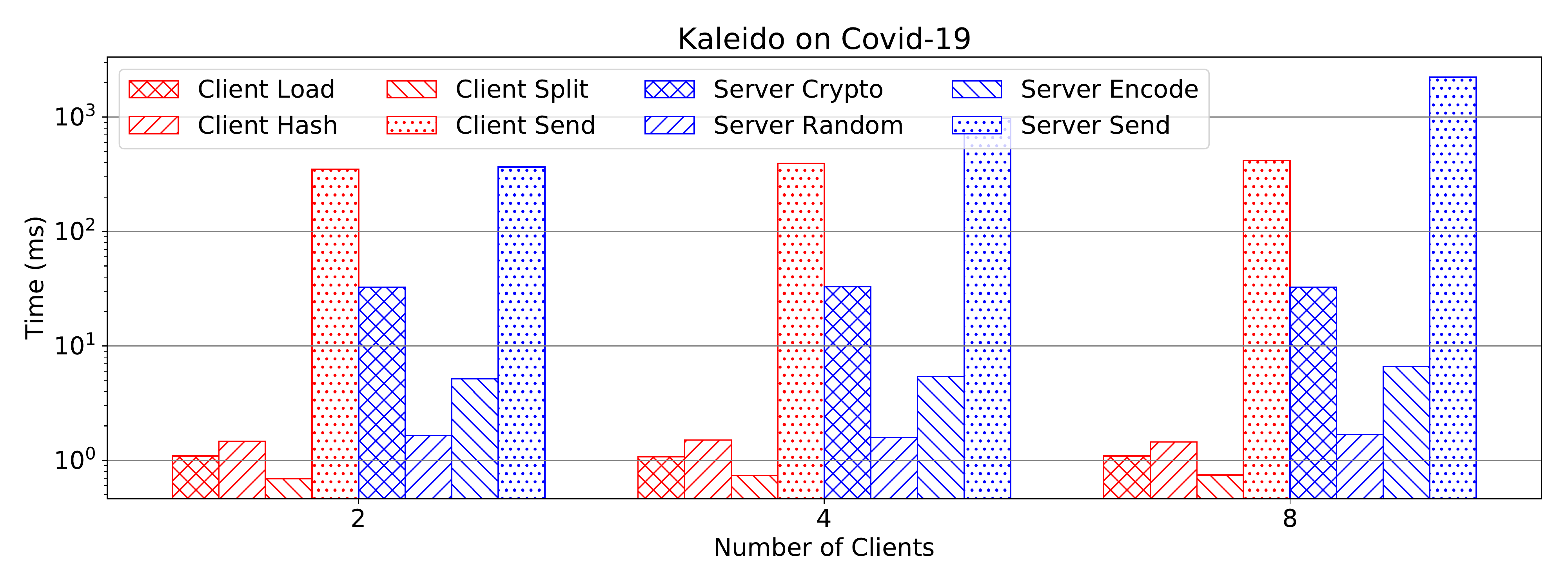}
  \caption{Kaleido performance on Covid-19 patient record~\cite{covid19data}}
  \label{fig:covid19}
\end{figure}

Figure~\ref{fig:covid19} reports the stage-wise performance of Kaleido and Prism on the COVID-19 data.
For clients, we observe that the communication from clients to serves dominates the cost.
For servers, again, the communication from servers to clients takes the majority of the overall cost,
and the portion increases when more clients are involved.
As a result, although the cryptographic cost introduced by Kaleido is significant from the computational perspective,
the real bottleneck of the entire system lies in communication.
In fact, the computational cost of the cryptographic component can be further reduced (not explored in this work). 

\paragraph{Bitcoin}

\begin{figure}[!t]
  \centering
  \includegraphics[width=\figscale\linewidth]{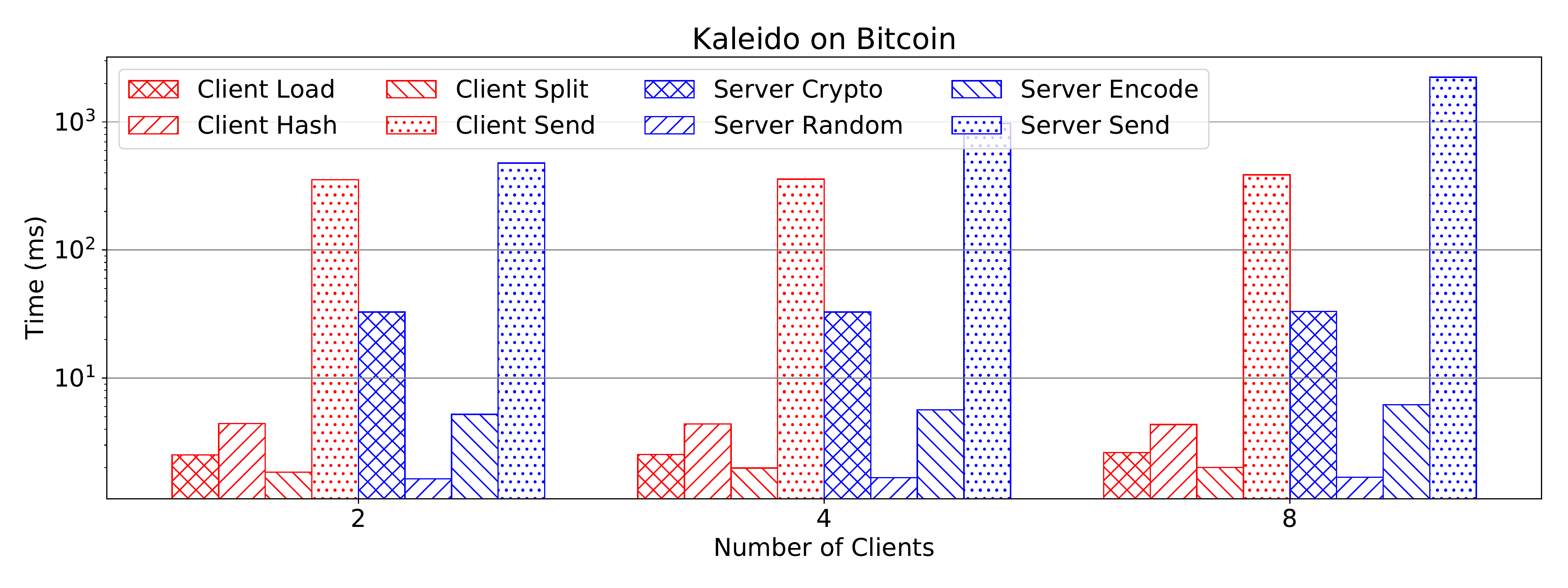}
  \caption{Kaleido performance on Bitcoin trade history~\cite{bitcoin_trade}}
  \label{fig:bitcoin}
\end{figure}

Figure~\ref{fig:bitcoin} reports the performance of the Bitcoin trade history.
We observe a similar pattern to the COVID-19 data.
Even more evidently, the numbers show that server communication dominates the overall cost on the server side:
When working with only two clients, 
the communication time of the servers takes 10x more time than the cryptographic component (i.e., AES~\cite{aes}).
These results reaffirm the practicality of the proposed Kaleido scheme:
Although the cost is significantly more than the naive random method,
the cost is an order of magnitude lower than the communication time.

\paragraph{Human Gene \#38 Sequence}

\begin{figure}[!t]
  \centering
  \includegraphics[width=\figscale\linewidth]{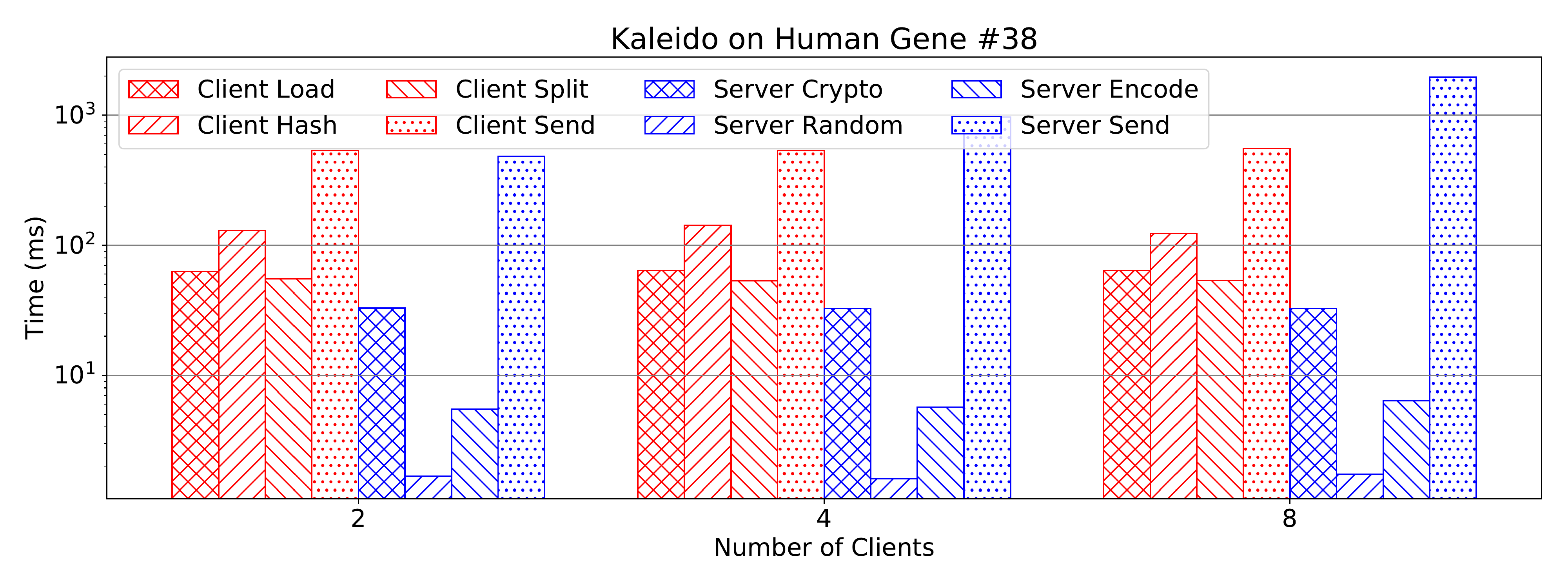}
  \caption{Kaleido performance on human gene \#38 sequence~\cite{hg_data}}
  \label{fig:hg38}
\end{figure}

Figure~\ref{fig:hg38} reports the performance on a larger data set---over 30,000 human gene \#38 sequence.
We observe that the server exhibits roughly the same performance as the previous two data sets;
however, the clients are more occupied this time.
This is understandable as the clients in this case must work on more data items regarding their local databases.

\section{Related Work }

This section reviews three other important techniques for ensuring the confidentiality of outsourced databases.
All of these techniques are orthogonal to the secret-share-based schemes (e.g., Prism, Kaleido).

\subsection{Encrypted Storage} 
The database instance from the cloud vendor is considered as storage of encrypted data and the client is responsible for nontrivial queries.
This solution is viable only if (i) the relations touched on by the query are small enough that the network overhead of transmitting those relations is acceptable, and (ii) the user has the capability (both computation and storage) to execute the query locally.
We stress that this solution might defeat the purpose of outsourcing the database service to the cloud.
    
\subsection{Encrypted Tuples} 

Every tuple of the original relation $R$ is encrypted into a ciphertext that is stored in column $T$ of a new relation $R^s$.
For each attribute $A_i$ in $R$, there is a corresponding attribute $A_i^s$ in $R^s$, whose value is the index of $R.A_i$.
The index is usually assigned by a random integer based on some partitioning criteria and can be retrieved with the metadata stored on the \textit{client}, i.e., the user's local node.
As a result, the schema stored at the cloud provider is $R^s(T, A_1^s, \dots, A_i^s, \dots)$.
When the user submits a query $Q$, the client splits $Q$ into two subqueries $Q_s$ and $Q_c$.
$Q_s$ serves as a filter to eliminate those unqualified tuples based on the indices in $R^s$ and transmits the qualified tuples (in ciphertexts) to the client.
$Q_c$ then ensures that those false-positive tuples are eliminated after the encrypted tuples are decrypted using the secret key presumably stored on the client.
This approach involves both the client (i.e., the user) and the server (i.e., the cloud provider) when completing a query, often referred to as \textit{information hiding} approaches~\cite{hhaci_sigmod02}.
    
\subsection{Encrypted Fields} 

The third approach aims to minimize the involvement of clients when processing the query over the encrypted data stored at the cloud vendor.
The idea is to encrypt the relations at a finer granularity---each attribute of a relation is separately encrypted.
The key challenge of this approach lies in its expressiveness,
e.g., how to apply arithmetic or string operations over the encrypted fields.
While fully homomorphic encryption (FHE)~\cite{cgentry_stoc09} can support a large set of computing problems,
the performance of current FHE implementations cannot meet the requirements of practical database systems~\cite{arx_vldb19,popa2011cryptdb}.
An alternative solution is a partially homomorphic encryption (PHE) schemes~\cite{ppail_eurocrypt99,elgamal_tit85},
which are orders of magnitude faster than FHE but only support a single algebraic operation.
Traditional PHE schemes are designed for public-key (asymmetric) encryption,
which is desirable for straightforward key distribution over insecure channels but significantly more expensive than secret-key (symmetric) encryption.
However, in the context of outsourced databases, 
the user usually serves as both the sender and the receiver and there is no need to distribute the key.
To this end, symmetric (partially) homomorphic encryption, was proposed~\cite{symmetria_vldb20,apapa_osdi16}.

% \paragraph{Secure Multi-Party Computation (MPC). }
% In addition to HE-based methods,
% another widely-used technique for data privacy is secure \textit{multi-party computation} (MPC),
% which originated from~\cite{ayao_focs82} and has been mostly built upon oblivious transfer~\cite{ogold_stoc87,mkeller_ccs16}, threshold homomorphic encryption~\cite{rcramer_eurocrypt01,idamgard_crypto03}, and secret sharing~\cite{ashamir_cacm79,trabin_stoc89}.
% MPC has been applied in multiple machine learning frameworks,
% such as DeepSecure~\cite{deepsecure}, SecureML~\cite{secureml}, and ABY~\cite{aby}.

\section{Conclusion and Future Work}

This paper first demonstrates a security vulnerability of a state-of-the-art PSI scheme, Prism, in outsourced databases.
The paper then proposes a series of new primitives, namely Kaleido, to extend Prism such that the new PSI scheme is semantically secure.
Both the correctness and security are formally proven,
and the intuition is explained with a running example.
A system prototype of Kaleido is implemented and deployed to two servers and 32 database owners hosted at CloudLab~\cite{cloudlab}.
Extensive experiments on the TPC-H benchmark and three real-world applications confirm the effectiveness of the proposed Kaleido schemes.

Our future work is two-fold.
First, we will leverage the underlying multiple/many CPU cores to parallelize the costly cryptographic operations in Kaleido.
The challenge lies in how to ensure the multiple (non-colluding) servers do not communicate and yet stay synchronized regarding the randomized group generators.
Second, we will adopt some compression schemes to reduce the vectors,
generated by both the clients and the servers,
such that the communication cost can be reduced.
The challenge of the second research direction lies in the balance between the computational cost for compressing and decompressing the vectors and the communication saving out of the compressed vectors.

% -------------------------------------------------------------------------------
\bibliographystyle{plain}
\bibliography{ref_new}

% \newpage
% \include{sigmod23_revision_request}

\end{document}